\documentclass[%
 aip,
 amsmath,amssymb,
 reprint,%
]{revtex4-2}

\usepackage{graphicx}
\usepackage{dcolumn}
\usepackage{bm}
\usepackage{float}

\usepackage[utf8]{inputenc}
\usepackage[T1]{fontenc}
\usepackage{mathptmx}
\usepackage{etoolbox}

\makeatletter
\def\@email#1#2{%
 \endgroup
 \patchcmd{\titleblock@produce}
  {\frontmatter@RRAPformat}
  {\frontmatter@RRAPformat{\produce@RRAP{*#1\href{mailto:#2}{#2}}}\frontmatter@RRAPformat}
  {}{}
}%
\makeatother
\begin{document}

\preprint{AIP/123-QED} 

\title[SWDMPM]{Spin wave diffraction model for perpendicularly magnetized films} %
\author{V. Vlaminck}
 \email{vincent.vlaminck@imt-atlantique.fr}
 \altaffiliation[Also at ]{Lab-STICC - UMR 6285 CNRS, Technopole Brest-Iroise CS83818, 29238 Brest Cedex 03}
\author{L. Temdie}%
 \altaffiliation[Also at ]{Lab-STICC - UMR 6285 CNRS, Technopole Brest-Iroise CS83818, 29238 Brest Cedex 03}
 \author{V. Castel}%
 \altaffiliation[Also at ]{Lab-STICC - UMR 6285 CNRS, Technopole Brest-Iroise CS83818, 29238 Brest Cedex 03}

\affiliation{ 
IMT- Atlantique, Dpt. MO, Technopole Brest-Iroise CS83818, 29238 Brest Cedex 03
}%
\author{M.B. Jungfleisch}
\affiliation{University of Delaware, Dept. of Physics and Astronomy, Newark, Delaware, 19716, USA 
}%

\author{D. Stoeffler}
\author{Y. Henry}
\author{M. Bailleul}
\affiliation{%
IPCMS - UMR 7504 CNRS
Institut de Physique et Chimie des Matériaux de Strasbourg, \\
Université de Strasbourg, 23 rue du Loess, BP 43, 67034 Strasbourg Cedex 2, France
}%

\date{\today}

\begin{abstract}
We present a near-field diffraction model for spin waves in perpendicularly magnetized films applicable in any geometries of excitation fields. This model relies on Kalinikos-Slavin formalism to express the dynamic susceptibility tensor in k-space, and calculate the diffraction patterns via inverse 2D-Fourier transform of the response functions. We show an excellent quantitative agreement between our model and MuMax3 micro-magnetic simulations on two different geometries of antennas. Our method benchmarks spin wave diffraction in perpendicularly magnetized films, and is readily applicable for future designs of magnon beamforming and interferometric devices. 

\end{abstract}

\maketitle


\section{INTRODUCTION} 
The growing field of magnonic focuses on the implementation of elementary magnetic excitations called spin waves -or their quanta magnons- for unconventional electronic applications \cite{Chumak2015,ChumakBook2019,Barman2021}. The peculiar wave nature of magnons along with their long coherent lengths constitute a formidable ground for new methods of data processing and computing \cite{Chumak2017,Chumak2022}, where information could be encoded in both the phase and the amplitude of the spin wave. Namely, spin waves are being studied as the building blocks of novel wave computing methods such as spin wave logic \cite{Khitun2010,Chumak2014,Talmelli2019}, holographic memory \cite{Khitun2013,Gertz2014,Khitun2015}, and neuromorphic computing \cite{Grollier2020,Papp2021}, which are all interference-based techniques.\\ 
Among these ideas, several works have also opened different strategies for shaping and manipulating spin wave beams. On one hand, a particular effort focused on applying the concepts of optics to grasp the reflection and refraction of spin waves at interfaces \cite{Gruszecki2014,Stigloher2016,Gruszecki2017,Gräfe2020}. On the other hand, the focusing or diffracting of spin wave beams in continuous layers have been demonstrated either by designing constricted microwave antennas \cite{Gruszecki2016,Körner2017,Loayza2018,Divinskiy2020,Bertelli2020} or also by using magnetic textures \cite{Albisetti2020}. Additionally, the steering of so-called caustic beams originating from a point-like source of anisotropic magnetostatic modes holds promising properties for integrated magnonic logic networks \cite{Heussner2020}. Ultimately, all of these examples of magnons beamforming can be understood in terms of near-field diffraction of an excitation source with specific spatial distribution.\\
Most generally, the mapping of these diffraction patterns require micromagnetic simulations, which come at a substantial cost in time and resources, and therefore restrict the scope of exploration. Surprisingly, an optically-inspired approach, which is far less demanding than finite element simulations, has remained underrated to tackle these near-field diffraction effects. In the initial work of Loayza et. al \cite{Loayza2018}, they adopt Huygens-Fresnel principles to map the spin wave amplitude from summing punctual circular waves over a rectangular slit. Although it successfully describes the focused emission of spin wave beams from a constricted coplanar wave guide (CPW), this approach lacked of quantitative meaning, and is not applicable for curved geometries. \\
In this communication, we present a model for mapping the near-field interferences of spin waves in film magnetized perpendicular-to-plane readily applicable for arbitrary shapes of antennas. We then present a comparative study between micromagnetic simulations and our model for two different geometries of excitation field: (i) a straight constriction in a coplanar wave guide (CPW), and (ii) a quarter-circular CPW.  
  
\begin{figure}[H]
\centering
\includegraphics[width=62mm]{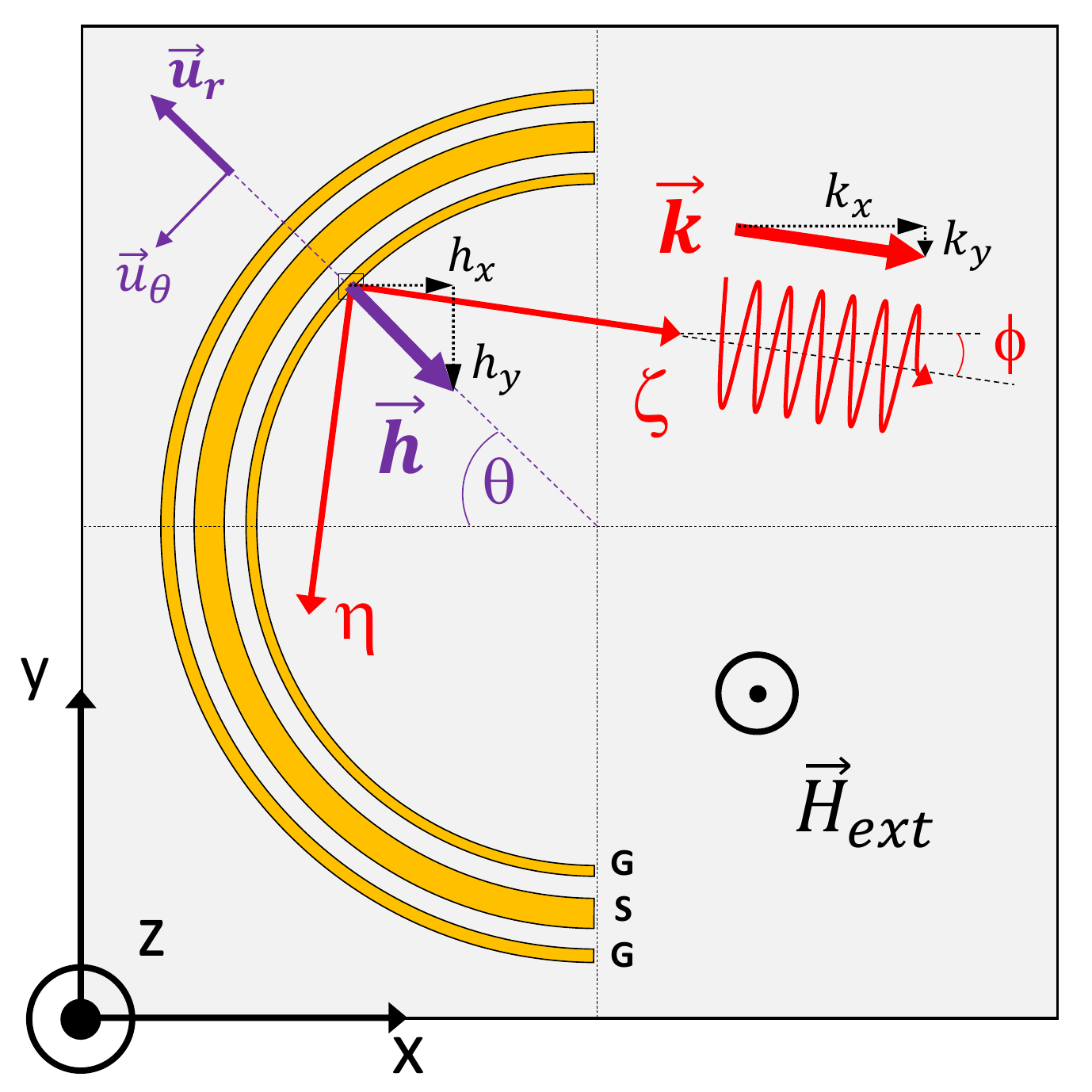}
\caption{\label{fig1} Geometry for the near-field diffraction model showing both frames of reference ($x$,$y$,$z$), and ($z$,$\eta$,$\zeta$)}
\end{figure}

\section{NEAR-FIELD DIFFRACTION MODEL}
Our model consists in extending Eq.(30) of Kalinikos's seminal work \cite{Kalinikos1981}, which only applies to an infinitely long and straight antenna, to more general cases of spatially resolved antennas in the 2D-space. This requires to establish the linear response functions $\chi(k_x,k_y)$ to an excitation $\vec{h}(k_x,k_y)$, in order to calculate the dynamic amplitude of the magnetization $\vec{m}(x,y)=(m_x,m_y)$ from its inverse Fourier transform:
\begin{equation}
\vec{m}(x,y,t)=e^{i\omega t}\iint_{-\infty}^{+\infty}dk_xdk_y \chi(k_x,k_y) \vec{h}(k_x,k_y)  e^{-i(k_xx+k_yy)}
\label{eq1}
\end{equation}

\begin{figure}[t]
\includegraphics[width=90mm]{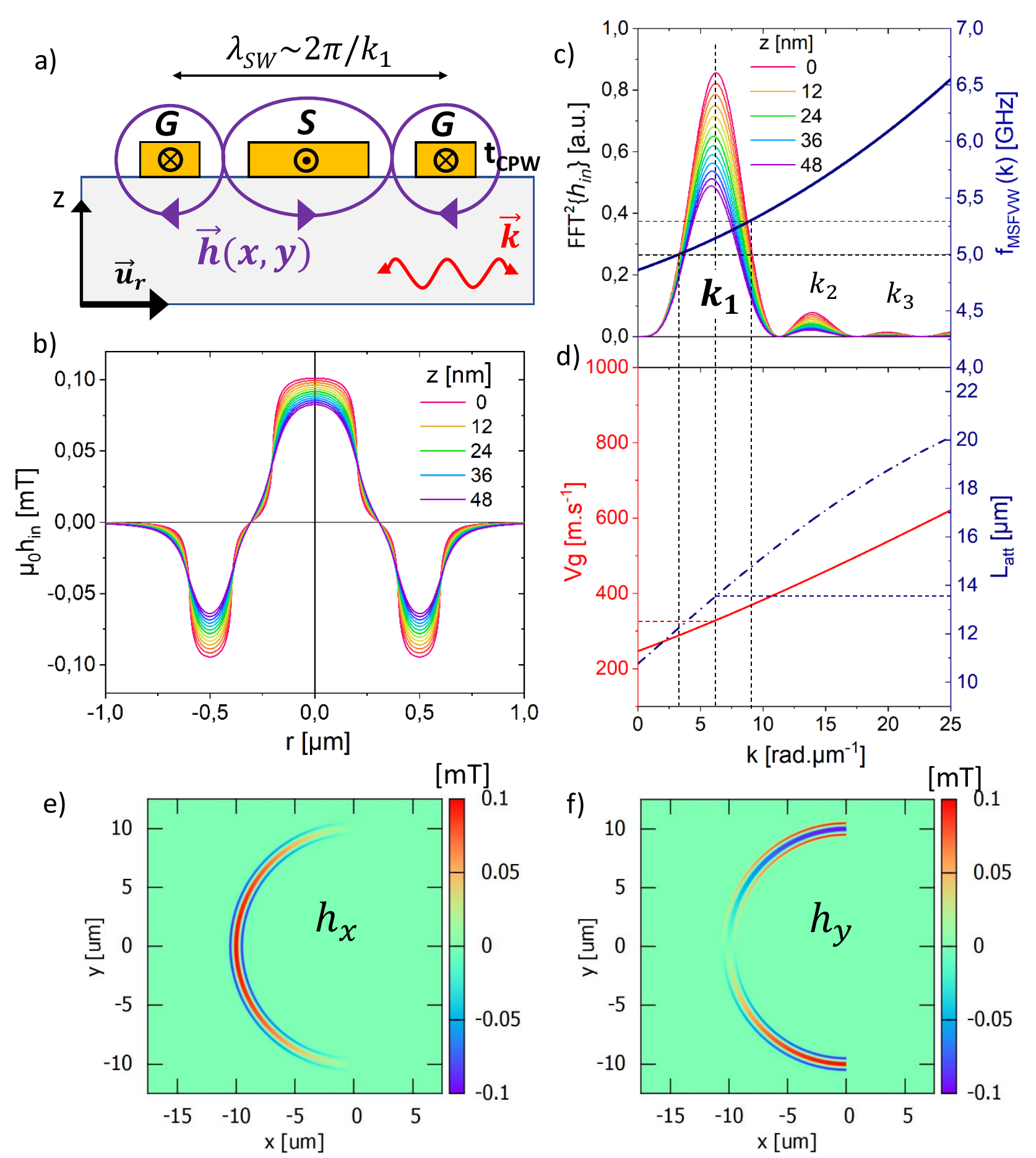}
\caption{\label{fig2}(a) Sketch of the excitation field for a CPW section. (b) Thickness dependence of the in-plane $h$ for a CPW of dimensions $S=400\,$nm, $G=200\,$nm, and spaced by $200\,$nm. (c) Fourier transform of the in-plane $h_{exc}$ with dispersion relation for $\mu_0H_{ext}=308\,$mT. (d) Group velocity and attenuation length for the same field, $f=5.15\,$GHz and $\alpha=7.5\,10^{-4}$. Mapping in the ($x$,$y$) plane of the components (e) $h_x$, and (f) $h_y$ for a semi-circular CPW.  }
\end{figure}

We will describe now the important steps used to establish the susceptibility tensor in reciprocal space $\chi(k_x,k_y)$. Following Kalinikos-Slavin Green functions approach \cite{Kalinikos1986}, we define two frames of reference, as sketched in Fig.1. 
We first define the lab frame ($x$,$y$,$z$) to linearize Landau-Lifschitz Gilbert (LLG) equation, which we orientate in such a manner that an experimental detection method would be sensitive to one of the planar component of $\vec{m}(x,y)$, and also that the equilibrium direction coincides with the $z$-direction. 
The second frame ($z$,$\eta$,$\zeta$) is related to a spin "wavelet" mode ($k_n$) propagating in the $\zeta$-direction, which forms an angle $\phi$ with the $x$-direction. The use of this second frame is particularly convenient to define the dynamic dipolar tensor from the Green function tensor $\mathbb{G}_{z,\eta,\zeta}(z-z')$ \cite{Guslienko2011}: 
\begin{equation}
\mathbb{N}_{ij}^{(n_1,n_2)}=-\iint_{-t/2}^{t/2}dzdz' \; m_{j,n_2}\:\mathbb{G}_{ij}(z-z')\: m_{i,n_1}
\label{Nij_n1n2}
\end{equation}
where the spin wave modes ($m_{x,n}$,$m_{y,n}$) are written in the set of plane waves propagating in the $\zeta$-direction with a standing wave profile along the film thickness $t$ in accordance with the chosen pinning conditions at the film surfaces:
\begin{equation}
\vec{m}(z,\zeta,k)=\sum_{n} \left(\begin{array}{c}m_{x,n}(z)\\m_{y,n}(z)\end{array}\right) e^{-ik\zeta}
\label{eq4}
\end{equation}
In the present study, we will restrict ourselves to the fundamental mode ($n=0$), assume no coupling with higher order modes, and consider unpinned conditions ($\left( \frac{\partial{\vec{m}}}{\partial{z}}\right)_{\pm t/2}=\vec{0}$) at both top and bottom surfaces. This leads to a uniform profile of the spin wave amplitude across the thickness, and the dynamic dipolar tensor in the ($x$,$y$) frame can be written as:
\begin{equation}
\mathbb{N}_{ij}^{(0,0)}=\begin{pmatrix} P \cos^2(\phi) & \frac{P}{2}\sin(2\phi) \\ \frac{P}{2}\sin(2\phi) & P \sin^2(\phi) \end{pmatrix}
\label{eq5}
\end{equation}
where $P=1-\dfrac{1-e^{-kt}}{kt}$ is obtained from Eq.(\ref{Nij_n1n2}) for $n_1=n_2=0$, and for $k=\sqrt(k_x^2+k_y^2)$. \\
In the lab frame, the linearized LLG equation accounting for the exchange and dipolar dynamic fields can be written as:
\begin{equation}
\vec{h}=\left(\begin{pmatrix} 0 & -1 \\ 1 & 0 \end{pmatrix}i\frac{\omega}{\omega_M}+(\Omega_k+i\alpha\frac{\omega}{\omega_M})\mathbb{I} + \mathbb{N}_{ij}^{(0,0)}\right)\vec{m}
\label{eqLLG}
\end{equation}
where $\vec{h}$ is the excitation field from an antenna with arbitrary spatial distribution in the 2D plane, $\omega_M=\gamma \mu_0 M_s$, $\Omega_k= \frac{\gamma \mu_0 H_{equ}}{\omega_M}+\Lambda^2 k^2$ with $\Lambda^2=\dfrac{2A_x}{\mu_0Ms^2}$ the exchange length, $\mathbb{I}$ the identity matrix, and $\mathbb{N}_{ij}$ the dynamic demagnetizing tensor.
Finally, the susceptibility tensor in reciprocal space $\chi(k_x,k_y)$ is obtained by inverting Eq.(\ref{eqLLG}): 
\begin{equation}
\chi_{ij}=\dfrac{1}{|\chi^{-1}|}\begin{pmatrix} \Omega_k + i\alpha\dfrac{\omega}{\omega_M} P \sin^2(\phi) & i\alpha\dfrac{\omega}{\omega_M}-\frac{P}{2}\sin(2\phi) \\ -i\alpha\dfrac{\omega}{\omega_M}-\frac{P}{2}\sin(2\phi) & \Omega_k + i\alpha\dfrac{\omega}{\omega_M} P \cos^2(\phi)) \end{pmatrix}
\label{eq6}
\end{equation}
with the determinant of the system $|\chi^{-1}|=\frac{1}{\omega_M^2}(\omega_{res}^2-\omega^2+i\alpha\omega\omega_M(2\Omega_k+P))$, and \\
$\omega_{res}^2=\Omega_k\omega_M^2(\Omega_k+P)$ is the dispersion relation for the n=0 mode, which is plotted in Fig.\ref{fig2} for an external field $\mu_0H_{ext}=308\,$mT. \\

The remaining task consists in defining properly the excitation field components ($h_x$,$h_y$) in the ($x$,$y$) plane for the curved geometry of interest. For this purpose, we made the approximation that the excitation field is perpendicular to the section of the antenna at all point as sketched in Fig.2-(a), and use the expression of the Oersted field produced by a wire of rectangular section \cite{Chumakov2007}. This assumption is rather fair considering the quasi-TEM nature of modes in coplanar waveguides. We plotted in Fig.2-(b) the spatial distribution of the in-plane excitation field at various depths throughout the magnetic film ranging from 0 to 50\,nm, for a relatively narrow CPW of central line $S$=400\,nm, $G$=200\,nm, spaced by 200\,nm and with thickness $t_{CPW}$=40\,nm, which mainly couples to magnons of wavelength $\lambda_{SW}\approx 1\mu$m. We chose these dimensions for all of our simulations as they correspond to typical nano-antennas fabricated on a YIG film. For such narrow waveguides, the field distribution presents some inhomogeneities across the thickness, as well as slightly faster decrease of the field produced by the ground lines. This leads to a small dependence of the excitation range across the film thickness as shown with the Fourier transform $h^2(k)$ in Fig.2-(c), which not only decreases in amplitude but also slightly shifts towards lower wavevector as z increases. Obviously, our 2D model only considers the coupling to an excitation at a fixed height. Therefore, it will be even more relevant for thinner films (up to 50nm thickness). Otherwise, one would need to compute several height in order to build an average response. We also provide an example of the $h_x$ and $h_y$ field components for a semi-circular CPW, which then have to be Fourier transformed and multiplied by the susceptibility tensor defined in k-space in Eq.(\ref{eq6}).\\

\section{RESULTS AND DISCUSSION}
We chose two geometries to compare our Near-Field Diffraction model (NFD) with the corresponding micromagnetic simulations performed with MuMax3 \cite{MuMax3}: (i) a straight piece of CPW, (ii) a quarter-circular CPW. \\
All of our simulations used a 30-nm thin YIG film, with saturation magnetization $\mu_0 M_s$=198\,mT, uniaxial out-of-plane anisotropy $\mu_0 H_{K_u}$=62\,mT (or $K_u$=4880\,J.$m^{-3)}$), an exchange constant $A_{exch}$=4\,pJ.$m^{-1}$, a gyromagnetic ratio $\gamma$=28.26\,GHz.$T^{-1}$, and a Gilbert damping constant $\alpha$=7.5\,$10^{-4}$. These values correspond to a real sputtered YIG film that was previously characterized \cite{Jungfleisch2017,Loayza2018}, which explains the relatively high value of the damping. All the micromagnetic simulations were performed with an external out-of plane static field of $\mu_0 H_{ext}$=308\,mT, and at 8 different frequencies ranging from 5.00\,GHz to 5.35\,GHz.\\ 
The excitation field was defined from straight heaviside functions of 0.1mT amplitudes constant across the thickness with the same lateral dimensions defined above ($S$=400\,nm, $G$=200\,nm, spaced by $200$\,nm) in order to remain in the linear regime response. The slight inhomogeneities of the excitation field mentionned above were not considered for the present study. In order to optimize calculation time with the micromagnetic simulations, we used a $20*30\mu m^2$ window discretized in 10*10 nm2 cells surrounded by a 500\,nm thin absorbing contour within which the  Gilbert damping increases gradually to 1 in order to avoid reflections. Also, a single cell was defined across the entire thickness as we focused on the n=0 uniform PSSW mode. All of the simulations presented below will display the mapping of the amplitude of the x-component of the dynamic magnetization ($|m_x|(x,y)$). In the case of MuMax3 simulations, we adopted a $sin(2\pi\,f\,t)$ excitation, and recorded the spatial and temporal evolution of the x-component of the dynamic magnetization $m_x(x,y,t) = |m_x(x,y)| sin(2\pi\,f\,t+\Phi(x,y))$ for each pixel over a full period $T=1/f$ after the steady state had been reached, which typically takes about 200*T to occur. In the case of the NFD simulations, we represented the modulus quantity $|m_x|(x,y)=\sqrt{Re(m_x)^2+Im(m_x)^2}$ of the spatial term of Eq.(\ref{eq1}).

\begin{figure}[t]
\includegraphics[width=90mm]{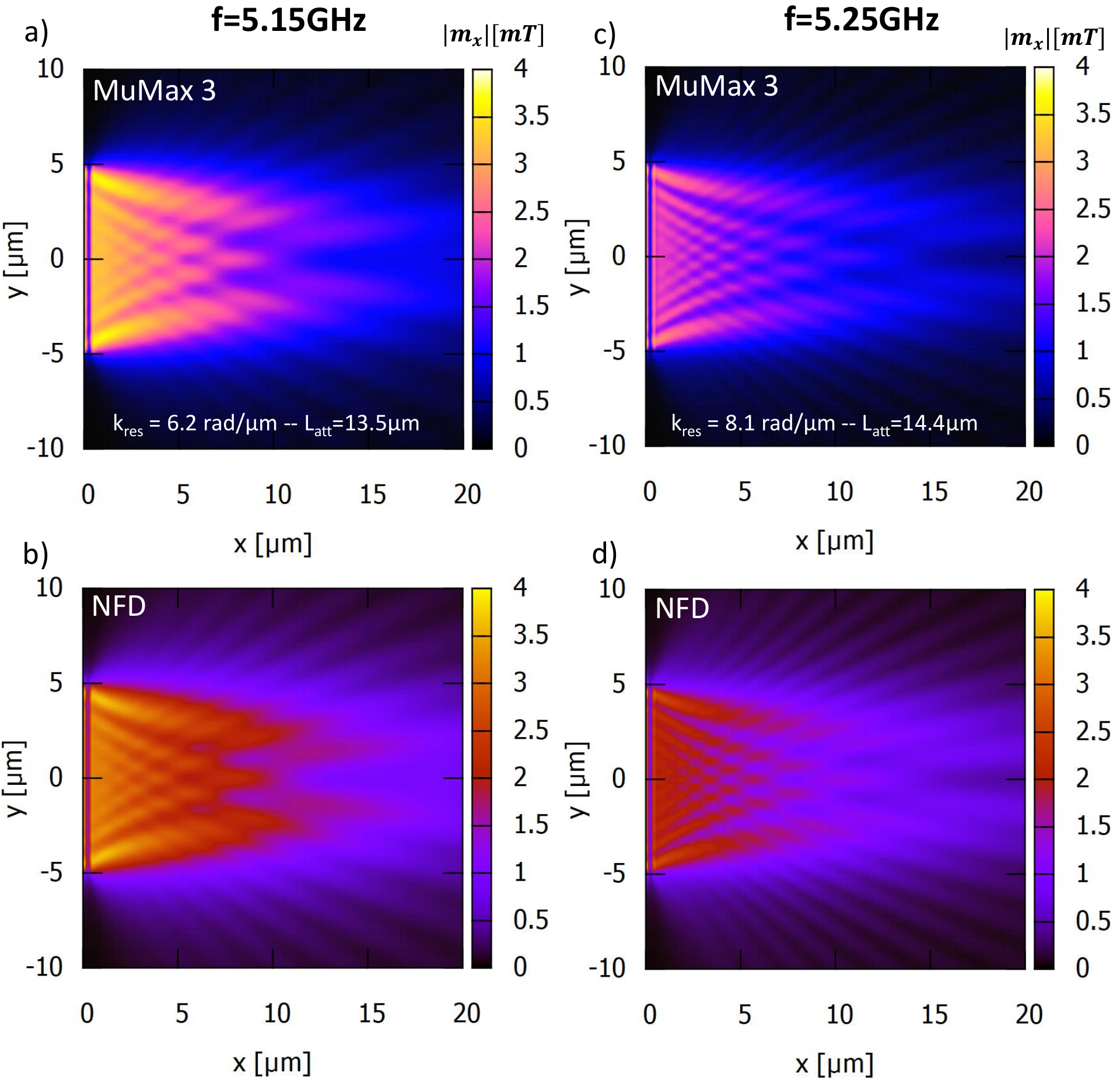}
\caption{\label{fig:epsart} Amplitude $|m_x|(x,y)$ of the x-component of the dynamic magnetization for a 10µm segment of CPW respectively at f=5.15\,GHz, and f=5.25\,GHz. (a) and (c) are the MuMax3 simulations, and (b) and (d) are the corresponding near-field diffraction simulations.}
\end{figure}

\subsubsection{Spin Wave Beams in CPW segment}
For the first geometry, we used an isolated straight CPW segment of 10$\mu$m length, and with the same lateral dimensions S, G, and $t_{CPW}$ defined above. Such an antenna can be seen as the equivalent of a single slit in optics, and was successfully crafted using abrupt transitions between the constriction and the rest of the CPW to excite spin wave beams \cite{Loayza2018}.\\
We present in Fig.3 the mapping obtained with both MuMax3 and our near-field diffraction model at two different frequencies: a first one at $5.15\,GHz$ which corresponds to the resonance frequency at the peak of the Fourier transform (namely $k_{res}\approx6.17\,rad.\mu m^{-1}$, see Fig. 2-c)), and a second one at $5.25GHz$ for which the coupling to the film occurs for slightly higher wavevector ($k_{res}\approx7.98\,rad.\mu m^{-1}$) and is therefore less efficient. The correspondence between MuMax3 and NFD simulations is absolutely remarkable in both qualitative and quantitative aspects of the diffraction patterns within the spin wave beam. Our model reproduces all the small features of the diffraction pattern. This confirms in particular that the focused emission of a spin wave beam from a sharply constricted CPW is solely due to its shape and its resulting near-field diffraction effect. On another note, the near-field/far-field limit, which is defined as the ratio $D^2/\lambda_{SW}\approx 100\,\mu m$, with $D$ being the antenna length and $\lambda_{SW}$ the magnon wavelength, will be mostly out-of sight in magnonic devices. Furthermore, the initial Fresnel diffraction model proposed by Loayza et al. \cite{Loayza2018}, which simply consist in summing punctual circular waves over a rectangular slit, already captures most of the features of the diffraction patterns. However, it can not easily provide a quantitative information on the magnetization amplitude, and it is not adapted to tackle curved geometries. \\
Finally, it is worth noting that just a single one of these mapping obtained from MuMax3 takes about 48 hours using a GTX1080Ti graphic card with 11\,Go of RAM, while the same mapping are obtained in less than a minute with the near-field diffraction approach on any laptop.

\subsubsection{Quarter-circular CPW}

\begin{figure}
\includegraphics[width=85mm]{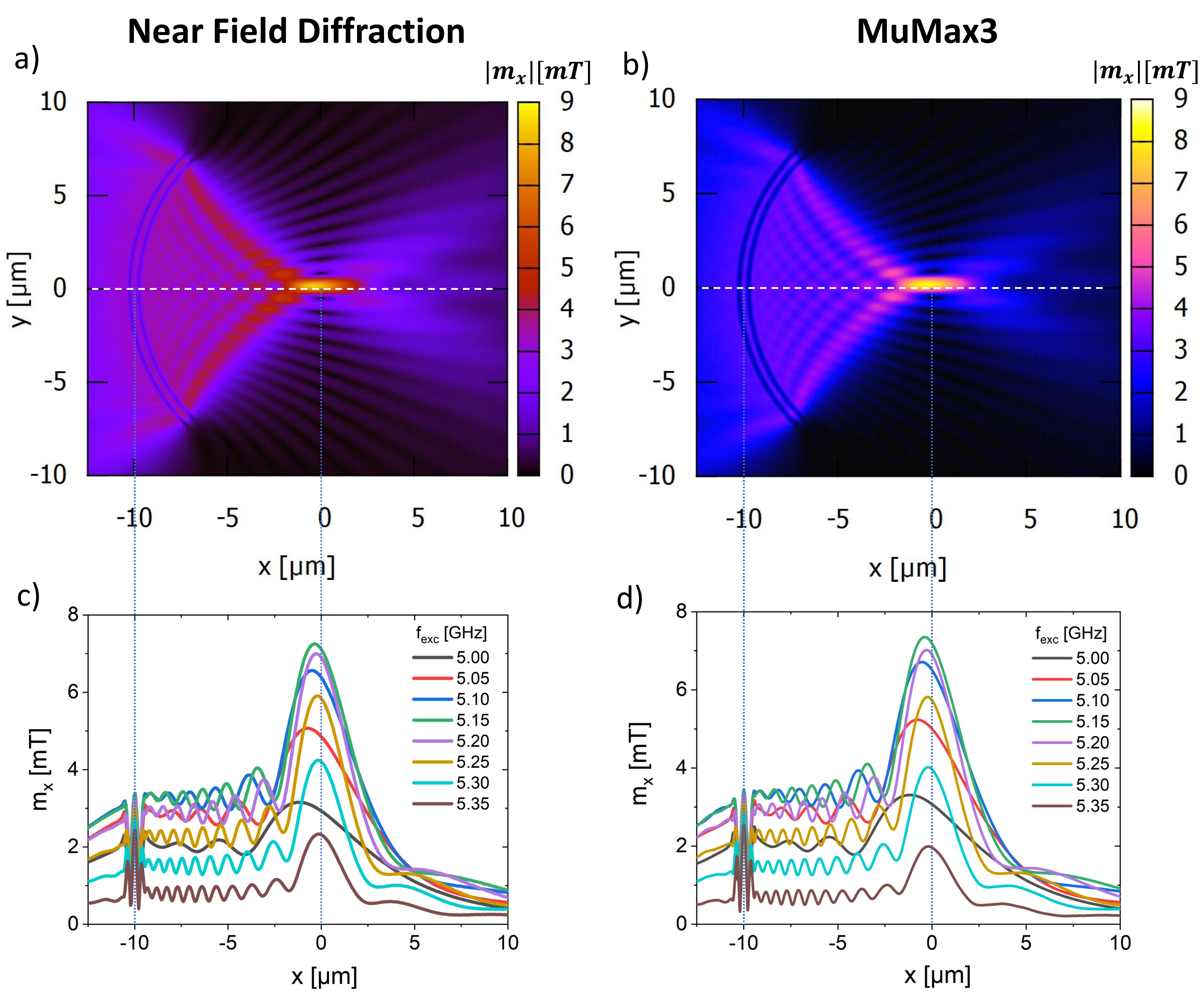}
\caption{\label{fig4} Diffraction pattern from a quarter-circular CPW of radius $R=10\,\mu m$ at f=5.15\,GHz obtained with (a) NFD model, and (b) MuMax3 simulations. Cross-section of $m_x$ at $y=0$ for various frequencies for (c) NFD, and (d) MuMax3.}
\end{figure}

We chose for our second geometry a quarter-circular CPW with radius $R$=10\,$\mu$m as a conclusive test for validating our model in curved field distributions. 
We did not consider any dephasing of the excitation field along the $\frac{\pi}{2}R$ antenna length as it remains far smaller than the electromagnetic wavelength in the wave guide, which are typically of the order of cm in the microwave spectrum. Here again, such an isolated circular antenna could be achieved with sufficiently abrupt transitions at its extremities.   
We present in Fig.4 the diffraction patterns at the peak frequency of 5.15\,GHz obtained with both NFD and MuMax3 simulations, as well as the cross-section of the amplitude $m_x$ at $y=0$ for 8 different frequencies ranging from 5.00 to 5.35 GHz. Once again, both diffraction pattern are nearly identical both qualitatively and quantitatively, thereby confirming the applicability of our approach in arbitrary geometries. \\
We observe that such a quarter-circular CPW acts as a "spin wave concentrator" as the amplitude near the focal point ($x=0$,$y=0$) is approximately doubled from its value under the antenna ($\tilde{m}_{x,max}=m_{x,max}/m_{x,ant}\approx 2.18$). Besides, this amplification factor can be also be anticipated using the simple Fresnel approach \cite{Loayza2018}, and considering each element of the antenna as a source of coherent circular waves. In that case the amplitude at the focal point normalized by the value at excitation gives $\tilde{m}_{x,max}=\frac{\pi}{2}\sqrt{R}\,exp(-R/L_{att})\approx 2.37$ using the corresponding attenuation length $L_{att}$ given in Fig.2-d).\\            
Nevertheless, a closer look at the two diffraction patterns shows that the maximum of amplitude does not happen right at the focal point but is slightly up-shifted in the y-direction, although the antennae was carefully designed symmetrically with respect to the y=0 line, and the out-of-plane spin wave modes are known to have isotropic properties. It is remarkable that both the micromagnetic simulations and the near-field diffraction simulations reveal this unexpected upshift, while the Fresnel's approach from Loayza et al.\cite{Loayza2018} does not show any asymmetry. We attribute this asymmetry to the difference in parity with respect to the y-direction between the two components $h_x$ and $h_y$ of the driving field. Namely, the driving field $h_y$ is odd in the y direction, while $h_x$ is even. Therefore, the spin wave response being the superposition of both components of the driving field, it will be neither even or odd with respect to the y-axis. This peculiarity could open up new perspectives for the exploration of non-reciprocal magnonic devices.

\section{CONCLUSION}

We presented a very efficient method to simulate magnons diffraction patterns in perpendicularly magnetized film applicable in any geometry of excitation field. Our approach reproduces quantitatively and 3000 times faster the corresponding micromagnetic simulations.
This model therefore constitutes a benchmarking tool to explore magnon beamforming as well as magnons interferometry at a much lower cost in time and resources, and it will allow to broaden the scope of study in field, frequency range, magnetic parameters, magnetic anisotropy, film thickness, shape and scale of antenna.
In essence, the same approach can and will be extended to in-plane modes, or even arbitrary orientation of external field.

\section{Supplementary Materials}
See Supplementary Materials for the finite element simulation of the microwave field of a circular antenna, the mapping of the difference between MuMax3 and near-field diffraction simulations, and the frequency dependence of the interference pattern for the first three peaks of the antenna.

\begin{acknowledgments}
We would like to thank the reviewer for providing an explanation on the asymmetry of the circular geometry. The authors acknowledge the financial support from the French National research agency (ANR) under the project \textit{MagFunc}, the Département du Finistère through the project \textit{SOSMAG}, as well as the High Performance Computing Center of the University of Strasbourg for supporting this work by providing access to computing resources. Part of the computing resources were funded by the Equipex Equip@Meso project (Programme Investissements d'Avenir) and the CPER Alsacalcul/Big Data.
\end{acknowledgments}

\section*{Data Availability Statement}

The data that support the findings of this study are available from the corresponding author upon reasonable request.











\end{document}